\documentclass[journal]{IEEEtran}
\usepackage{amsmath,amssymb,amsfonts, bbm}
\usepackage{graphicx}
\usepackage{cite}
\usepackage{bm}
\usepackage[caption=false, font=footnotesize]{subfig}
\usepackage{xcolor}

\newcommand{\Pe}{\acute{P}_e}
\newcommand{\Pbar}{\bar{P}_e}
\newcommand{\gbar}{\bar{\gamma}}
\newcommand{\Lerch}{\Phi}

\begin{document}
\bstctlcite{IEEEexample:BSTcontrol}

\font\myfont=cmr12 at 21pt

\title{\myfont Average Finite-Blocklength Packet Error Rate over Nakagami-$m$ Fading via a Logistic--Lerch Approximation}
\author{Aamir Mahmood,~\IEEEmembership{Senior~Member,~IEEE}%
\thanks{A. Mahmood is with the Department of Computer and Electrical Engineering, Mid Sweden University, 851~70 Sundsvall, Sweden (e-mail: aamir.mahmood@miun.se).}
\vspace{-22pt}
}
\maketitle

\begin{abstract}
Evaluating the average packet error rate (PER) of finite-blocklength (FBL) coded transmission over fading requires integrating the block-error waterfall, given by the normal approximation, against the fading distribution, which is intractable for general Nakagami-$m$ channels. This letter approximates the conditional waterfall by a slope-matched logistic function and shows that its Nakagami-$m$ average reduces to a single Lerch-transcendent term that interpolates between the FBL waterfall and the classical outage limit, with norming constants explicit in rate and blocklength. The closed form supports non-integer fading and, composed into an effective-capacity objective, yields a quality-of-service (QoS) aware rate-selection rule. It matches the normal-approximation integral to about 1\% uniformly in $m$ over the nominal FBL operating region, while outage and linearization baselines exceed several percent at high diversity.
\end{abstract}

\begin{IEEEkeywords}
Finite blocklength, PER, Nakagami-$m$ fading, Lerch transcendent, rate adaptation, URLLC.
\end{IEEEkeywords}

\section{Introduction}
\IEEEPARstart{S}{hort}-packet communication is central to machine-type and ultra-reliable low-latency communication (URLLC) services, where small payloads must meet stringent reliability and latency targets. Here, the long-codeword assumption no longer holds, and the packet error rate (PER) must be characterized in the finite-blocklength (FBL) regime and averaged over the fading distribution to reflect the link's actual reliability. However, accurate closed-form expressions for this fading-averaged FBL PER are difficult to obtain. A useful precedent is the uncoded case: an extreme value theory (EVT)-based method approximates the additive white Gaussian noise (AWGN) PER by a Gumbel distribution function, which is easily integrable over Nakagami-$m$ fading and yields an accurate closed-form average PER across a wide range of SNR and packet lengths~\cite{evt2016}. That method, however, relies on the per-bit product structure of the uncoded PER, $\Pe(\gamma)=1-(1-b_e(\gamma))^N$, with $b_e(\gamma)$ the bit-error probability at signal-to-noise ratio (SNR) $\gamma$ and $N$ the packet length in bits, which does not hold for coded transmission.

 For coded transmission, the conditional (AWGN) block-error probability is instead characterized by the normal approximation~\cite{ppv2010}, a sharp sigmoidal waterfall in the SNR. Averaging this waterfall over fading is the quantity of interest, as it underlies reliability assessment and resource allocation in fading links. More refined approximations of the FBL error exist, notably saddlepoint methods~\cite{saddlepoint}, which approximate the underlying coding bounds more accurately than the normal approximation, but yield no closed-form fading average. Closed-form results exist for Rayleigh fading~\cite{yang2014,makki2014}, but accurate ones for general Nakagami-$m$ remain scarce because the normal-approximation integral is intractable. The existing tractable approaches, most commonly a piecewise-linear approximation of the error curve~\cite{makki2014} or its coarser outage (step) limit~\cite{yang2014}, are tight only at low fading diversity, and degrade as $m$ increases, precisely the high-reliability URLLC regime. For Nakagami-$m$ fading specifically, FBL performance has been characterized through closed-form upper and lower bounds on the average block-error rate, together with capacity-outage and symbol-error quantities~\cite{zhang2019icc}, and through energy- and error-rate analyses for URLLC~\cite{zhang2019globecom}. These works provide bounds or optimization-oriented characterizations of the fading-averaged error. In contrast, this letter targets a direct closed-form approximation of the fading-averaged PER itself, obtained by averaging a logistic surrogate of the waterfall rather than by bounding the error.
 
In this letter, we propose a new closed-form approximation of the average PER for coded short-packet transmission over Nakagami-$m$ fading. Following the EVT-based approach of~\cite{evt2016} for uncoded schemes, we approximate the AWGN error curve by a tractable function before averaging over fading, but with a logistic rather than a Gumbel form. Our key contributions are: (i) a shared approximation strategy that replaces the AWGN error curve by a regime-specific tractable sigmoid before averaging, Gumbel for uncoded and a slope-matched logistic for the coded FBL waterfall; (ii)  a single-term closed-form average PER over Nakagami-$m$ via the Lerch transcendent, valid for all real $m>0$ (including sub-Rayleigh $1/2\le m<1$) and reducing to the classical outage probability as $N\to\infty$; (iii) closed-form norming constants of the logistic function obtained directly from the normal approximation; and (iv) a QoS-aware rate-selection rule obtained by composing the closed form into a cross-layer effective-capacity objective.

\section{System Model and Sigmoid Approximation}
\label{sec:model}
We consider a coded packet of blocklength $N$ channel uses and rate $R$ (in bits/channel use) transmitted over a block-fading channel with instantaneous received SNR $\gamma$ and average SNR $\gbar$. Over one block, the fading gain is constant, so conditioned on a realization $\gamma$, the channel is AWGN, with conditional block-error probability $\Pe(\gamma)$. The quantity of interest is its expectation over the fading distribution, which, by the law of total probability, is the average PER experienced over the link, where one packet corresponds to one coded block, defined as
\begin{equation}
\Pbar(\gbar)=\int_0^\infty \Pe(\gamma)\,p(\gamma;\gbar)\,d\gamma,
\label{eq:avg}
\end{equation}
where $p(\gamma;\gbar)$ is the SNR density under Nakagami-$m$ fading, i.e., a gamma density with shape (fading) parameter $m$,
\begin{equation}
p(\gamma;\gbar)=\frac{m^m \gamma^{m-1}}{\gbar^m \Gamma(m)}\exp\!\Big(\!-\frac{m\gamma}{\gbar}\Big),\quad \gamma\ge 0 .
\label{eq:gamma}
\end{equation}
In the FBL regime, the conditional error $\Pe(\gamma)$ is accurately given by the normal approximation~\cite{ppv2010}
\begin{equation}
\Pe(\gamma)\approx Q\!\left(\frac{C(\gamma)-R}{\sqrt{V(\gamma)/N}}\right),
\label{eq:ppv}
\end{equation}
valid for $N\gtrsim 100$~\cite{ppv2010}, where $C(\gamma)=\log_2(1+\gamma)$ is the AWGN capacity, $Q(x)=\frac{1}{\sqrt{2\pi}}\int_x^\infty e^{-t^2/2}\,dt$ is the Gaussian $Q$-function, and
$V(\gamma)=\big(1-\frac{1}{(1+\gamma)^2}\big)(\log_2 e)^2$
is the (complex) AWGN channel dispersion~\cite{ppv2010,yang2014}.
 
The evaluation in~\eqref{eq:avg} is thus a standard two-stage average over fading; the FBL feature is only the shape of the inner conditional error~\eqref{eq:ppv}, a sigmoidal waterfall in $\gamma$, which the following approximation exploits.

Our approach rests on a strategy shared by both regimes: approximate the conditional error curve by a tractable sigmoid, then average over fading. For the uncoded PER, the maximum-of-$N$ product structure places $\Pe(\gamma)$ in the domain of attraction of the Gumbel law by EVT~\cite{evt2016}, so the approximating sigmoid $G(\gamma)=1-\exp(-\exp(-(\gamma-a)/b))$, whose average~\eqref{eq:avg} is tractable, is derived rather than postulated. The coded FBL curve~\eqref{eq:ppv} has no such maximum-of-$N$ structure, so EVT does not supply a derived sigmoid here. Nonetheless, the same strategy of approximating the error curve before averaging carries over, because the waterfall is a steep, nearly symmetric sigmoid, accurately matched by the logistic function
\begin{equation}
\Pe(\gamma)\approx \frac{1}{1+\exp\!\big((\gamma-a)/b\big)}.
\label{eq:logistic}
\end{equation}
with $a$ a location parameter (the waterfall center) and $b>0$ a scale parameter (its width) as the norming constants. \textcolor{black}{Near the waterfall center, the argument of~\eqref{eq:ppv} crosses zero and is locally linear in $\gamma$, so the waterfall approaches a probit (Gaussian-CDF) sigmoid, increasingly so as $N$ grows since $b\propto1/\sqrt{N}$ (Sec.~\ref{sec:constants}). The logistic is a well-known probit surrogate: a minimax fit of the standard normal CDF, $\Phi_{\mathrm N}(x)\approx1/(1+e^{-1.702x})$, has maximum error $0.0095$~\cite{bowling2009}, while our slope matching yields the scaling $2\sqrt{2/\pi}\approx1.596$.} 
Among common symmetric sigmoids (probit/$\mathrm{erf}$, $\tanh$, generalized logistic), the logistic is preferred because its complementary form $\Pe=z/(1+z)$ with $z=e^{-(\gamma-a)/b}$ has a Laplace-domain structure well suited to the fading average. Other choices, such as the probit/erf form, do not yield the same compact single-term fading average, whereas the $\tanh$ form is equivalent to a reparameterized logistic.  The closed-form average of Sec.~\ref{sec:closed} holds for any such $a,b$, and their explicit values, obtained directly from the normal approximation, are given in Sec.~\ref{sec:constants}. There, the logistic form proves the more accurate match across essentially all of the practically relevant $(N,R)$ space, while the Gumbel form remains preferable only in the ultra-short, low-rate corner.

\section{Closed-Form Average PER}
\label{sec:closed}
Evaluating $\Pbar(\gbar)$ in~\eqref{eq:avg} is difficult because the FBL curve~\eqref{eq:ppv} involves the $Q$-function of a nonlinear argument in $\gamma$. In what follows, we show that the logistic approximation~\eqref{eq:logistic} renders the average over Nakagami-$m$ fading in a compact closed form.

\emph{Proposition 1:} For the logistic approximation~\eqref{eq:logistic}, the average FBL PER over Nakagami-$m$ block fading is
\begin{equation}
\;\Pbar(\gbar)\approx \left(s b\right)^m\, e^{a/b}\, \Lerch \Big(-e^{a/b},\,m,\,sb+1\Big),
\label{eq:main}
\end{equation}
where $s=m/\gbar$ and $\Lerch(\zeta,\nu,\alpha)$ denotes the Lerch transcendent~\cite[\S25.14]{dlmf}.
 
\emph{Proof:} Substituting the logistic approximation~\eqref{eq:logistic} and the Nakagami-$m$ density~\eqref{eq:gamma} into the average~\eqref{eq:avg} gives
\begin{equation}
\Pbar(\gbar)=\frac{s^m}{\Gamma(m)}\int_0^\infty \frac{\gamma^{m-1}e^{-s\gamma}}{1+e^{(\gamma-a)/b}}\,d\gamma,\quad s=\frac{m}{\gbar}.
\label{eq:int}
\end{equation}
Substituting $x=\gamma/b$ and multiplying numerator and denominator of the integrand by $e^{-x}$ gives
\begin{equation}
\Pbar(\gbar)=\frac{(sb)^m e^{a/b}}{\Gamma(m)}\int_0^\infty \frac{x^{m-1}e^{-(sb+1)x}}{1-(-e^{a/b})e^{-x}}\,dx ,
\label{eq:sub}
\end{equation}
which matches the Lerch integral representation~\cite[\S25.14]{dlmf} (valid for $\nu>0$, $\alpha>0$)
\begin{equation}
\Lerch(\zeta,\nu,\alpha)=\frac{1}{\Gamma(\nu)}\int_0^\infty\frac{x^{\nu-1}e^{-\alpha x}}{1-\zeta e^{-x}}\,dx,
\label{eq:lerchrep}
\end{equation}
with $\zeta=-e^{a/b}$, $\nu=m$, and $\alpha=sb+1$. Although the defining series for $\Lerch$ diverges for our argument $\zeta=-e^{a/b}$ (as $|\zeta|>1$), the representation~\eqref{eq:lerchrep} provides its analytic continuation and is well defined and real-valued for $\zeta<0$. By~\eqref{eq:lerchrep}, the $\frac{1}{\Gamma(m)}$-weighted integral in~\eqref{eq:sub} is exactly $\Lerch(-e^{a/b},m,sb+1)$, leaving the prefactor $(sb)^m e^{a/b}$ and yielding~\eqref{eq:main}.  \hfill$\blacksquare$

Since~\eqref{eq:lerchrep} holds for any real $\nu=m>0$, the result is valid for gamma-distributed SNR of arbitrary, including non-integer, shape $m>0$. In classical Nakagami-$m$ envelope fading, this covers the full physical range $m\ge 1/2$, including the sub-Rayleigh interval $1/2\le m<1$ that integer-$m$ derivations exclude. This is practically relevant since $m$ is a continuous fading-severity parameter. In contrast to the uncoded (Gumbel) result, which requires a distinct combination of digamma and trigamma functions for each integer $m$~\cite{evt2016}, the logistic average~\eqref{eq:main} is a single special function holding for all $m$, including non-integer values. As a consistency check, in the limit where the waterfall sharpens ($b\to0$; equivalently $N\to\infty$, since $b\propto1/\sqrt{N}$ as shown in Sec.~\ref{sec:constants}), the logistic~\eqref{eq:logistic} tends to the step $\mathbbm{1}[\gamma<\gamma^\star]$ and~\eqref{eq:main} reduces to the classical Nakagami outage probability $\Pbar\to P(m,\,m\gamma^\star/\gbar)$, with $P(m,x)=\gamma(m,x)/\Gamma(m)$. The closed form thus interpolates between this outage limit as $N\!\rightarrow\!\infty$ and the finite-blocklength waterfall at moderate $N$.

\section{Norming Constants and Regime Selection}
\label{sec:constants}
We obtain the norming constants $(a,b)$ of the logistic approximation~\eqref{eq:logistic} by matching it to the FBL curve~\eqref{eq:ppv} in value and slope at the rate threshold $\gamma^\star=2^R-1$, where $C(\gamma^\star)=R$ and $\Pe(\gamma^\star)\approx \tfrac12$. Equating the logistic center value to $\tfrac12$ gives $a=\gamma^\star$. Differentiating~\eqref{eq:ppv} at $\gamma^\star$, where the $Q$-argument vanishes so that $Q'(0)=-1/\sqrt{2\pi}$, leaves only the numerator term, the slope $-C'(\gamma^\star)\sqrt{N/V(\gamma^\star)}/\sqrt{2\pi}$. Equating this to the logistic slope $-1/(4b)$ at $\gamma=a$ yields
\begin{equation}
a(R)=2^R-1,\quad b(R, N)=\sqrt{\frac{\pi}{2}}\,\frac{1}{2\,C'(\gamma^\star)}\sqrt{\frac{V(\gamma^\star)}{N}},
\label{eq:ab}
\end{equation}
with $C'(\gamma^\star)=\log_2 e/(1+\gamma^\star)$. Note $b\propto 1/\sqrt{N}$, in contrast to the $\log N$ scaling of the uncoded case. These closed-form constants render~\eqref{eq:main} fully self-contained. Because $b$ matches the slope at $\gamma^\star$, it gives a local fit; a fixed, modulation-independent rescaling $b\!\rightarrow\!0.94\,b$ yields the global least-squares fit over the waterfall\textcolor{black}{; interestingly, this is numerically consistent with the probit connection of Sec.~\ref{sec:model}, since $2\sqrt{2/\pi}/1.702\approx0.938$~\cite{bowling2009}}. We use the analytic $b$ throughout.

\begin{figure}[!t]
    \centering
  \subfloat[$N=100$ ]{%
       \includegraphics[width=0.48\linewidth]{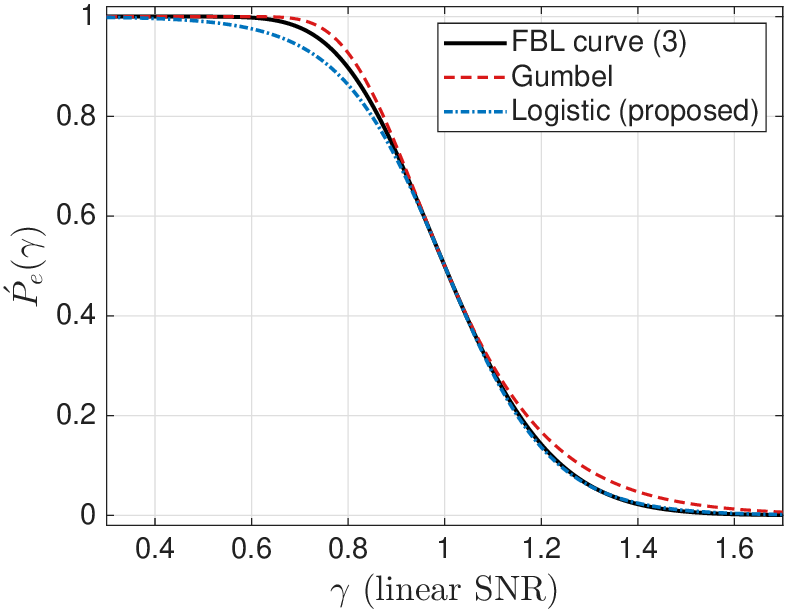}\label{fig:overlay-a}}\hfill
  \subfloat[$N=500$]{%
        \includegraphics[width=0.48\linewidth]{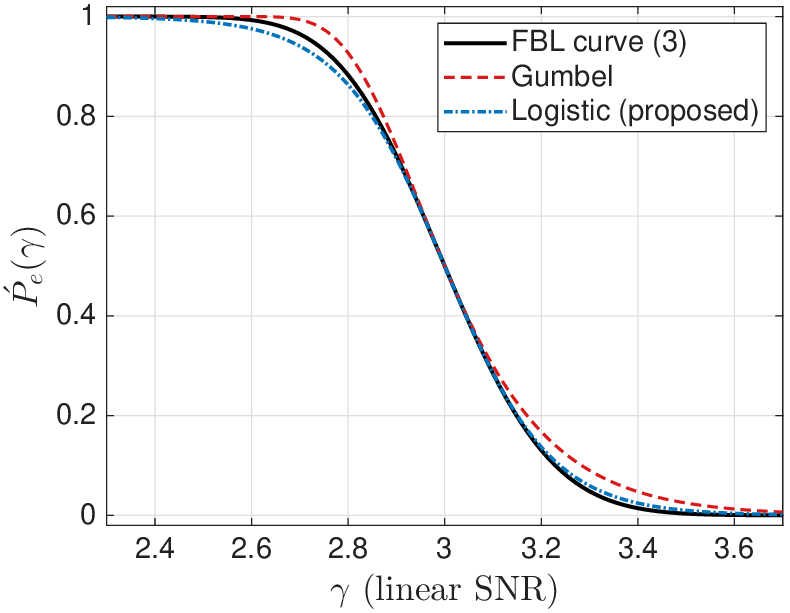}\label{fig:overlay-b}}              
  \caption{\textcolor{black}{Logistic and Gumbel approximations against the FBL error curve~\eqref{eq:ppv} at a fixed rate $R=1$.
  }}
  \label{fig:overlay}
\vspace{-14pt}  
\end{figure}

\emph{Regime selection:} Fig.~\ref{fig:overlay} overlays the logistic and Gumbel approximations on the FBL curve. For the Gumbel form $G(\gamma)=1-\exp(-\exp(-(\gamma-a_g)/b_g))$, the constants are obtained analytically by matching value and slope to~\eqref{eq:ppv} at the half-power point, giving $b_g=\ln2/(2s_0)$ and $a_g=\gamma^\star+b_g\ln(\ln2)$ with $s_0=C'(\gamma^\star)\sqrt{N/V(\gamma^\star)}/\sqrt{2\pi}$. The logistic form follows the symmetric waterfall closely, whereas the skewed Gumbel form deviates at the shoulders. \textcolor{black}{Crucially, the logistic fit improves with $N$: since $b\propto1/\sqrt{N}$, the waterfall sharpens and symmetrizes, favoring the symmetric logistic and penalizing the skewed Gumbel, reversing the trend of the uncoded case.} Taking the crossover where the two approximations attain equal root-mean-square (RMS) fit error over the waterfall interval $[\max(0,\gamma^\star\!-\!6b),\,\gamma^\star\!+\!6b]$, the logistic form is the better match for $N\gtrsim128$ across rates $R\gtrsim0.75$, while the Gumbel form is preferable only in the ultra-short, low-rate corner ($N\lesssim64$, low $R$). The two regimes are thus complementary.

\section{QoS-Aware Rate Selection}
\label{sec:opt}
Reliability at the physical layer shapes performance at higher layers, where delay and buffering impose statistical quality-of-service (QoS) constraints. A standard cross-layer metric is the effective capacity, which depends on the average error probability $\epsilon=\Pbar$. Since~\eqref{eq:main} provides $\epsilon$ in closed form, it composes directly into this metric without reintroducing the fading integral. Because $\epsilon$ is accurate (Sec.~\ref{sec:results}), the resulting optimal rate $R^\star$ tracks the normal-approximation optimum to well under $1\%$ across the tested $(N,m,\theta)$. Under a fixed-rate, on--off service model in which a packet of $NR$ bits is delivered with probability $1-\epsilon$ and otherwise lost, the effective capacity with QoS exponent $\theta$ is~\cite{shehab2019,qasmi2022}
\begin{equation}
R_E(\theta,N,R)=-\frac{1}{N\theta}\,\ln\!\Big(\epsilon+(1-\epsilon)\,e^{-\theta N R}\Big),
\label{eq:ec}
\end{equation}
with $\epsilon=\Pbar(\gbar;N,R,m)$, which correctly satisfies $R_E\to0$ as $\gbar\to0$ and $R_E\to R$ as $\gbar\to\infty$.

\begin{figure}[!t]
    \centering
  \subfloat[]{%
       \includegraphics[width=0.48\linewidth]{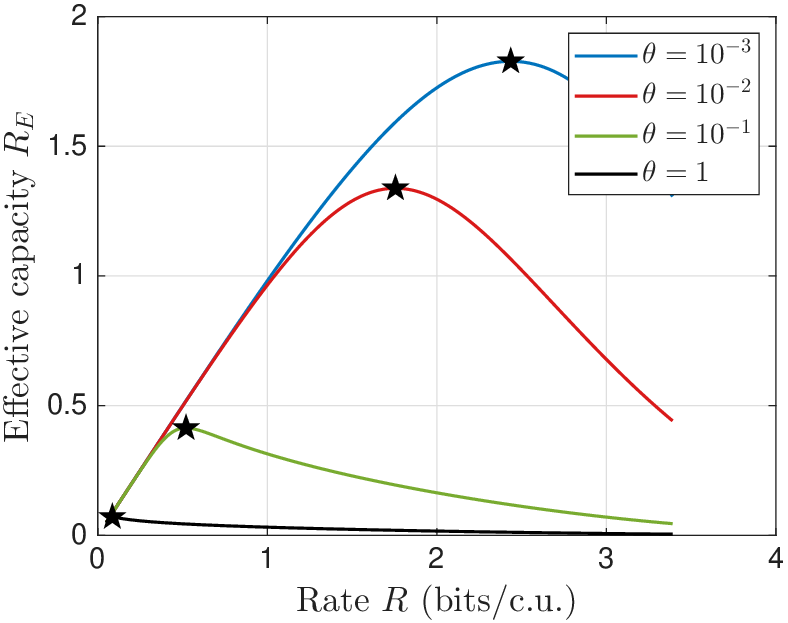}\label{fig:optrate-a}}\hfill
  \subfloat[]{%
        \includegraphics[width=0.48\linewidth]{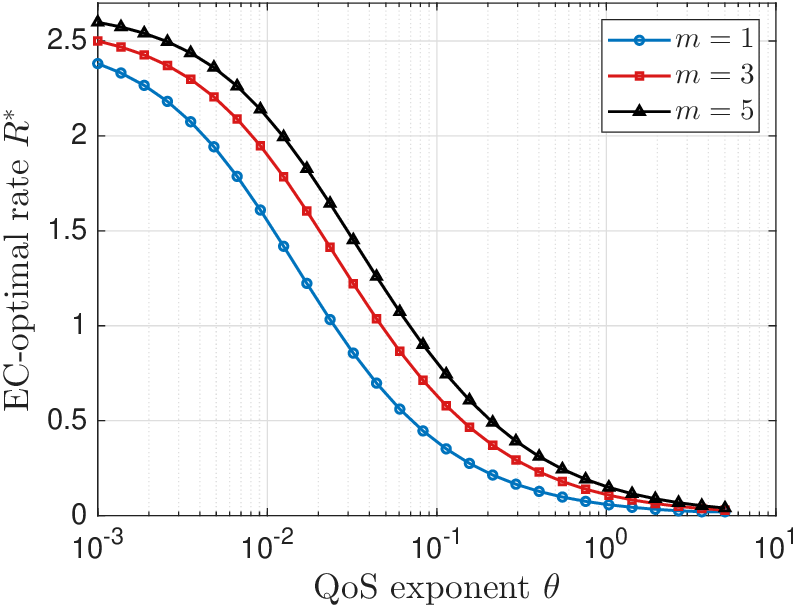}\label{fig:optrate-b}}     
  \vspace{-5pt}      
  \caption{QoS-aware rate selection via effective capacity: (a)~$R_E(R)$ in~\eqref{eq:ec} exhibits a single interior optimum $R^\star$ (markers) for $m=2$, $N=128$, $10$~dB; (b)~the optimal rate $R^\star$ versus QoS exponent $\theta$.}
  \label{fig:optrate}
\vspace{-14pt}  
\end{figure}

The transmission rate that maximizes $R_E$ trades the delivered payload $NR$ against reliability. To find it, we differentiate~\eqref{eq:ec} and set $\partial R_E/\partial R=0$.  The optimal rate $R^\star$ then satisfies the stationarity condition
\begin{equation}
\frac{\partial\epsilon}{\partial R}\big(1-e^{-\theta N R}\big)=\theta N\,(1-\epsilon)\,e^{-\theta N R},
\label{eq:stationary}
\end{equation}
which balances the marginal reliability loss $\partial\epsilon/\partial R$ against the marginal payload gain. Both $\epsilon$ and its derivative exist in closed form: $\epsilon$ from~\eqref{eq:main}, and $\partial\epsilon/\partial R$ by differentiating~\eqref{eq:main} via the elementary constants $a(R)$ and $b(R, N)$ of~\eqref{eq:ab}, which enter~\eqref{eq:main} through the prefactor $(sb)^m e^{a/b}$ and the first and third Lerch arguments $\zeta=-e^{a/b}$ and $\alpha=sb+1$. The two Lerch derivatives needed, $\partial_\zeta\Lerch(\zeta,\nu,\alpha)=\zeta^{-1}[\Lerch(\zeta,\nu-1,\alpha)-\alpha\Lerch(\zeta,\nu,\alpha)]$ and $\partial_\alpha\Lerch(\zeta,\nu,\alpha)=-\nu\Lerch(\zeta,\nu+1,\alpha)$, stay within the same special-function family as~\eqref{eq:main}, so $\partial\epsilon/\partial R$ is a closed-form combination of Lerch terms of orders $\nu-1,\nu,\nu+1$. Since $\epsilon$ is transcendental in $R$, \eqref{eq:stationary} has no elementary solution, but it is a single scalar equation that we solve by bisection. Across $m\in\{1,3,5\}$, SNR in $[5,20]$~dB, and $\theta\in[10^{-3},1]$, we find $R_E(R)$ to be concave with a single interior maximum (Fig.~\ref{fig:optrate-a}), so the root of~\eqref{eq:stationary} is unique. We do not claim concavity in general.

Fig.~\ref{fig:optrate-b} shows how $R^\star$ varies with the QoS exponent. As $\theta$ tightens from loose ($10^{-3}$) to stringent delay constraints, $R^\star$ falls sharply, as the link backs off its rate to meet the latency budget. For fixed $\theta$, $R^\star$ increases with the fading parameter $m$, since richer diversity sharpens the waterfall and permits operation closer to capacity. 

More fundamentally, because both $\epsilon$ and $\partial\epsilon/\partial R$ are explicit functions of $(N,R,m,\theta)$, the stationarity condition~\eqref{eq:stationary} exposes how the optimal rate responds to these parameters rather than hiding it in nested numerical integration. The monotone trends in Fig.~\ref{fig:optrate-b} read directly off this structure, as tighter delay constraints and lower diversity both narrow the feasible rate. Beyond this design insight, the closed form lowers the per-evaluation cost of QoS-aware rate optimization. In the linearization-based treatments~\cite{qasmi2022}, the averaged outage is a multi-term Gauss hypergeometric ${}_2F_1$ expression; here it is the single Lerch term~\eqref{eq:main}, whose derivative~\eqref{eq:stationary} stays within the same family, so each function and gradient evaluation in the rate search is lightweight. We solve only the rate sub-problem~\eqref{eq:stationary}; the joint power--rate energy-efficiency problem of~\cite{qasmi2022} is more general and outside our scope.

\section{Numerical Results and Discussion}
\label{sec:results}
We validate~\eqref{eq:main} against the normal-approximation average PER obtained by numerically integrating~\eqref{eq:ppv} over~\eqref{eq:gamma}. This numerical normal-approximation average serves as the reference, since~\eqref{eq:main} is a closed form for the logistic approximation rather than for the true coding error. All figures use the analytic norming constants of~\eqref{eq:ab} directly, without the optional $b\!\rightarrow\!0.94\,b$ recalibration.  

\begin{figure}[!t]
\centering
\includegraphics[width=0.85\columnwidth]{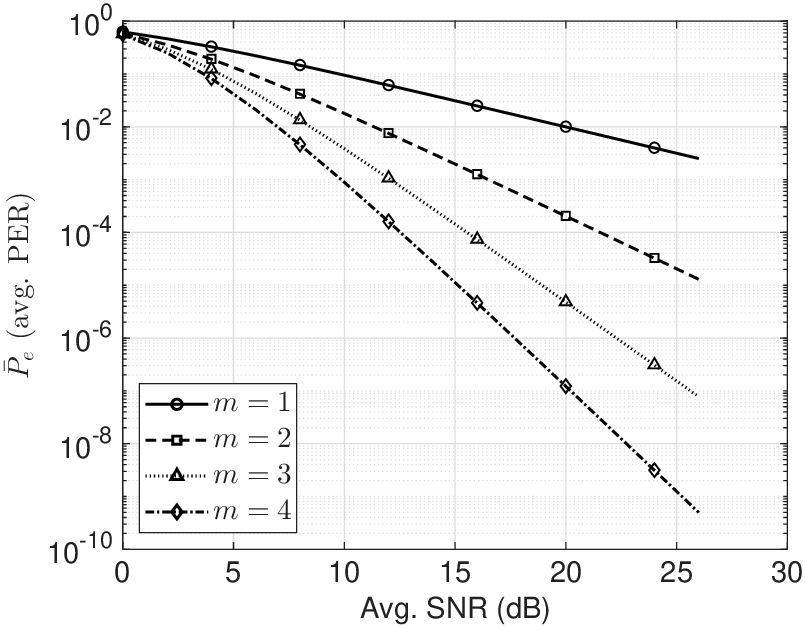}
\vspace{-10pt}
\caption{\textcolor{black}{Closed-form~\eqref{eq:main} (lines) versus normal-approximation average PER (markers) for Nakagami-$m$, $m=1,\dots,4$ ($N=128$, $R=1$).}}
\label{fig:summary}
\end{figure}

Fig.~\ref{fig:summary} plots~\eqref{eq:main} against the normal-approximation average PER for $m=1,\dots,4$. The closed form (lines) coincides with the numerical evaluation (markers) down to PER near $10^{-9}$, capturing the correct diversity slopes for every $m$. 
Table~\ref{tab:noninteger} reports the maximum relative error of~\eqref{eq:main} for $m\in\{0.7,1.5,2.5\}$, including sub-Rayleigh $m=0.7$. The error remains near $1\%$ throughout, confirming that the arbitrary-$m$ closed-form reduction holds in practice.
In evaluating~\eqref{eq:main}, the Lerch argument $\zeta=-e^{a/b}$ grows large as the waterfall sharpens (small $b$), so that $|\zeta|>1$. The closed form is nonetheless evaluated stably via the integral representation~\eqref{eq:lerchrep}, not the divergent series. \textcolor{black}{Moreover, after a simple rescaling, the normalized integrand of~\eqref{eq:sub} is of generalized Gauss--Laguerre form, so a fixed-order rule of a few tens of nodes sufficed at every tested operating point, whereas directly integrating~\eqref{eq:ppv} over~\eqref{eq:gamma} required adaptive quadrature with a few hundred evaluations.} The Lerch term is also a standard built-in (e.g., \texttt{LerchPhi} in Mathematica).
 
\begin{figure}[!t]
\centering
\includegraphics[width=0.9\columnwidth]{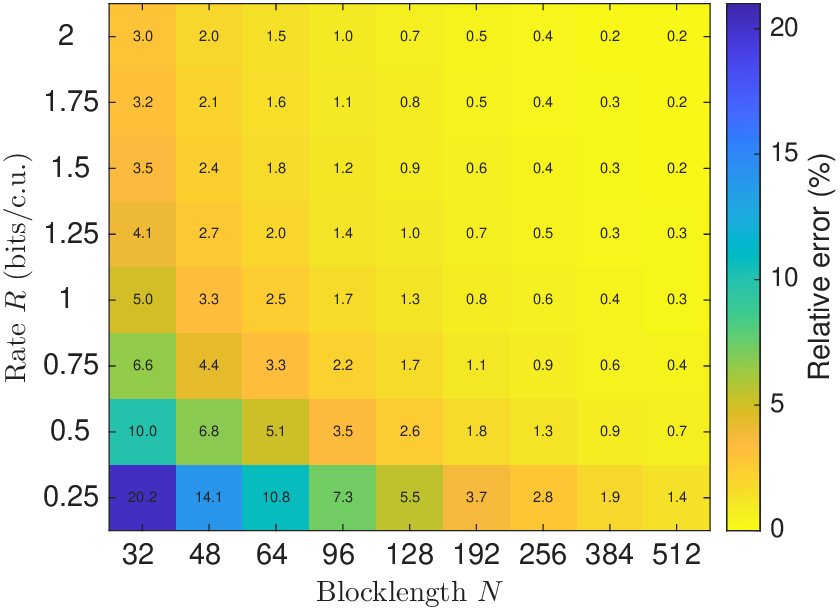}
\vspace{-6pt}
\caption{Relative error of the closed form~\eqref{eq:main} at $\Pbar\approx10^{-5}$ over the $(N,R)$ plane ($m=2$). Accuracy is below $1\%$ across most of the practically relevant region and degrades only for ultra-short, low-rate packets. The $N<100$ columns lie outside the normal approximation's nominal range (stress test).}
\label{fig:heatmap}
\vspace{-14pt}
\end{figure}

\begin{figure}[!t]
    \centering
  \subfloat[$m=1, N = 128, R=1$]{%
       \includegraphics[width=0.48\linewidth]{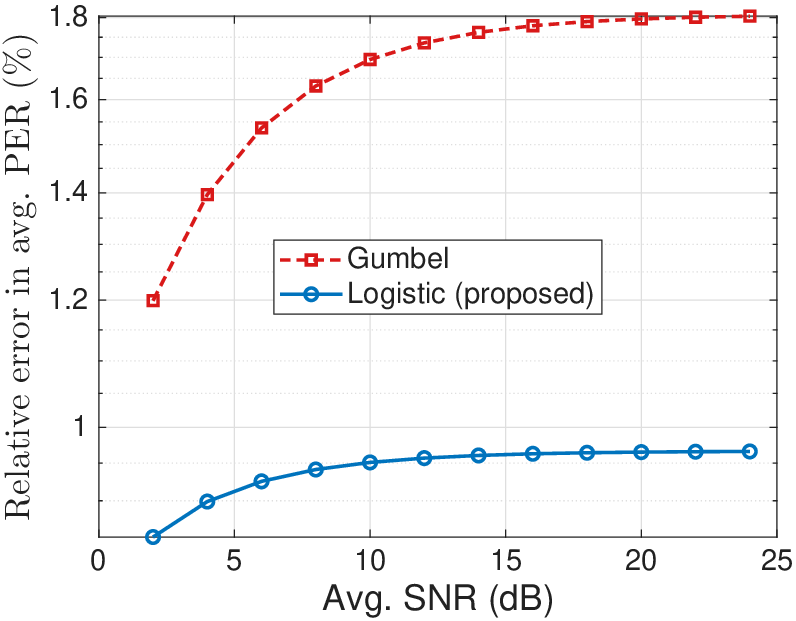}\label{fig:error-a}}\hfill
  \subfloat[$m=3, N = 128, R=1$ ]{%
        \includegraphics[width=0.48\linewidth]{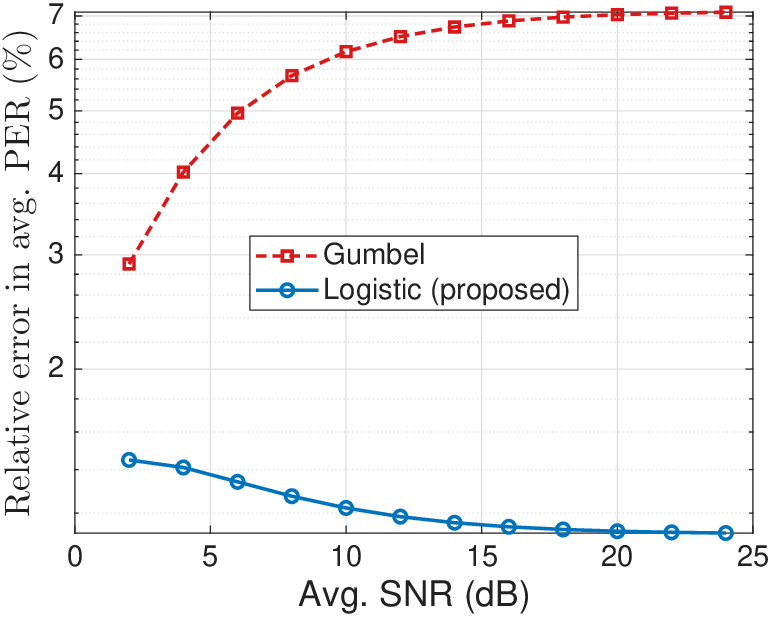}\label{fig:error-b}}              
  \caption{Relative error in average PER versus SNR for the logistic and Gumbel closed forms.}
  \label{fig:error}
\vspace{-14pt}
\end{figure}

\begin{figure}[!t]
    \centering
  \subfloat[Accuracy vs diversity]{%
       \includegraphics[width=0.9\linewidth]{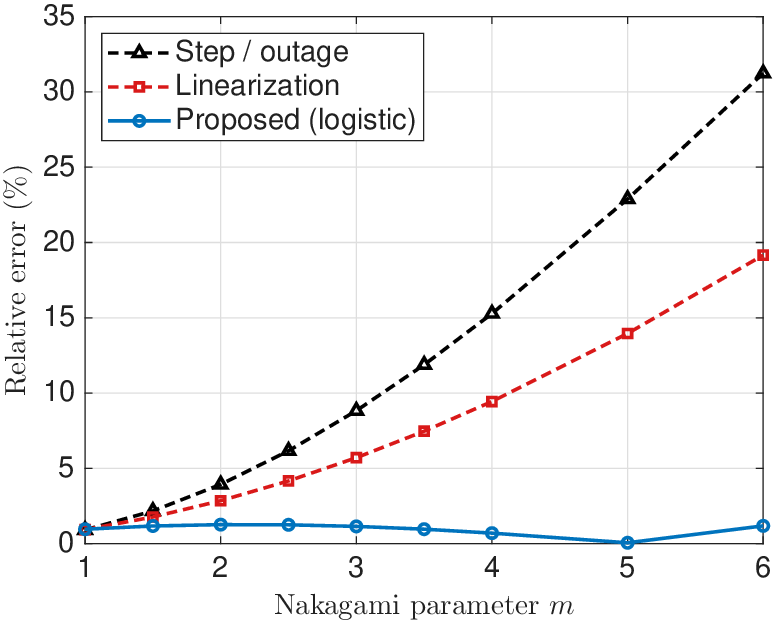}\label{fig:sota-a}}\\[-1pt]
  \subfloat[Accuracy vs SNR]{%
        \includegraphics[width=0.9\linewidth]{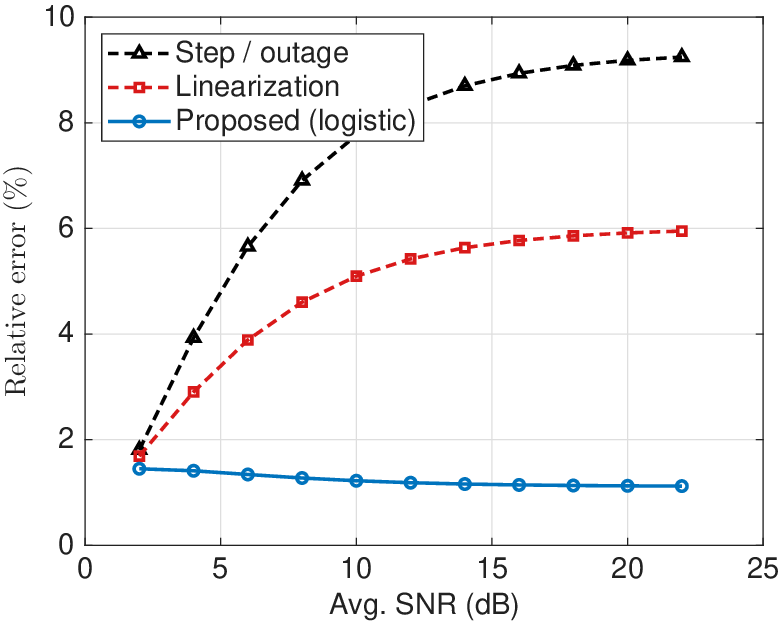}\label{fig:sota-b}}\\[-1pt]      
  \caption{Relative error of the proposed vs linearization~\cite{makki2014} and outage~\cite{yang2014} approximations. (a)~versus fading parameter $m$ (SNR=$15$~dB). (b)~versus SNR ($m=3$). Both at $N=128$, $R=1$.}
  \label{fig:sota}
\vspace{-12pt}  
\end{figure}

\begin{table}[!t]
\centering
\caption{Maximum relative error of the closed form~\eqref{eq:main} for non-integer fading parameters ($N=128$, $R=1$, SNR in $0$--$24$~dB).}
\label{tab:noninteger}
\begin{tabular}{cccc}
\hline
$m$ & $0.7$ (sub-Rayleigh) & $1.5$ & $2.5$\\
\hline
Max.\ rel.\ error & $0.75\%$ & $1.14\%$ & $1.36\%$\\
\hline
\end{tabular}
\vspace{-6pt}
\end{table}
The accuracy holds broadly across blocklength and rate. Fig.~\ref{fig:heatmap} maps the relative error of~\eqref{eq:main} over $N\in[32,512]$ and $R\in[0.25,2]$, with the SNR at each $(N,R)$ set to a target average PER of $\Pbar\approx10^{-5}$. It stays below $1\%$ across most of the plane, degrades gracefully toward the ultra-short, low-rate corner, and improves monotonically with both $N$ and $R$, consistent with the regime analysis of Sec.~\ref{sec:constants}. \textcolor{black}{In summary, the relative error in Fig.~\ref{fig:heatmap} stays below about $2\%$ within $N\gtrsim100$, $R\gtrsim0.75$, and below $1\%$ for $N\gtrsim192$, $R\gtrsim1$. The closed form is valid for arbitrary real $m>0$; tested integer and non-integer $m$ show comparable accuracy and weak target-PER sensitivity.}
The accuracy can also be bounded analytically: since the density in~\eqref{eq:avg} integrates to unity, the average error is controlled by the worst-case conditional fit between the normal-approximation and logistic curves
\begin{equation}
\big|\Pbar-\Pbar^{\mathrm{log}}\big|\le \sup_{\gamma\ge0}\big|\Pe^{\mathrm{NA}}(\gamma)-\Pe^{\mathrm{log}}(\gamma)\big|.
\label{eq:bound}
\end{equation}
This pointwise supremum is at most 6\% over the tested grid, yet the averaged error is several times smaller, near 1\%, because the fit error changes sign across the waterfall and partially cancels under the fading average. We stress that~\eqref{eq:bound} is a crude absolute-error bound. The relative errors, which are the figures of merit, are reported in Table~\ref{tab:noninteger} and Figs.~\ref{fig:heatmap}--\ref{fig:sota}.
 
With the accuracy of \eqref{eq:main} validated, we quantify the gain from the logistic choice. 
Fig.~\ref{fig:error} compares the logistic and Gumbel forms over SNR for $m=1$ and $m=3$ at $N=128$, $R=1$. The logistic approximation holds near $1\%$ across all SNRs and both fading parameters, whereas the Gumbel error grows from about $2\%$ at $m=1$ to roughly $7\%$ at $m=3$, so the logistic advantage widens with $m$, consistent with the regime analysis of Sec.~\ref{sec:constants}.
We finally compare~\eqref{eq:main} with two widely used closed-form approximations of the average FBL PER over fading: (i) the three-segment linearization of the error curve~\cite{makki2014}, recently applied to statistical-QoS and energy-efficiency analyses over Nakagami-$m$ fading~\cite{qasmi2022}, with slope $\mu=C'(\gamma^\star)\sqrt{N/V(\gamma^\star)}/\sqrt{2\pi}$ matched to~\eqref{eq:ppv} at the threshold $\gamma^\star$,
\begin{equation}
\Pe^{\mathrm{lin}}(\gamma)=
\begin{cases}
1, & \gamma<\gamma^\star-\tfrac{1}{2\mu},\\[2pt]
\tfrac12-\mu(\gamma-\gamma^\star), & |\gamma-\gamma^\star|\le\tfrac{1}{2\mu},\\[2pt]
0, & \gamma>\gamma^\star+\tfrac{1}{2\mu},
\end{cases}
\label{eq:lin}
\end{equation}
and (ii) the outage approximation $\Pe^{\mathrm{out}}(\gamma)=\mathbbm{1}[\gamma<\gamma^\star]$~\cite{yang2014}, which is exact in the asymptotic regime where quasi-static fading drives the FBL error toward the outage probability. Both baselines average in closed form over~\eqref{eq:gamma} through the regularized lower incomplete gamma function $P(m,x)=\gamma(m,x)/\Gamma(m)$, e.g., the outage form gives $\Pbar^{\mathrm{out}}=P\!\big(m,m\gamma^\star/\gbar\big)$, and the linearization adds the contribution of the middle segment, an elementary integral of $(\tfrac12-\mu(\gamma-\gamma^\star))$ against~\eqref{eq:gamma}. Fig.~\ref{fig:sota} reports the relative error of the proposed form and the two baselines. At $m=1$ the three are comparable, as the gentle waterfall is easy to approximate. As the fading parameter or SNR grows, however, the waterfall sharpens and shifts, and both baselines degrade substantially: the linearization to $\approx6\%$ and the outage form to $\approx9\%$ at $m=3$, rising to $\approx10\%$ and $\approx16\%$ at $m=4$, whereas the proposed approximation remains near $1\%$ throughout. 
\textcolor{black}{Against the closed-form baselines, the advantage of~\eqref{eq:main} is thus accuracy rather than special-function count (incomplete-gamma terms versus one Lerch term); the computational gain arises mainly against repeated numerical integration of the waterfall.}

A further class of methods approximates the Gaussian $Q$-function by exponential sums, notably the Chiani--Dardari--Simon (CDS) form $Q(x)\approx\tfrac{1}{12}e^{-x^2/2}+\tfrac14 e^{-2x^2/3}$~\cite{cds2003}. This closes over Nakagami-$m$ only for the classical argument $x\propto\sqrt{k\gamma}$, so that $e^{-x^2/2}$ is exponential in $\gamma$. The FBL argument $x(\gamma)=(C(\gamma)-R)\sqrt{N/V(\gamma)}$ is not of this form, so evaluating CDS over~\eqref{eq:gamma} still requires numerical integration. Meanwhile, forcing a closed form by linearizing the argument about $\gamma^\star$ yields error and parabolic-cylinder functions, with relative error growing from about $7\%$ at $m=1$ to nearly $30\%$ at $m=4$. Thus, CDS either requires numerical integration or, once expressed in closed form, is less accurate than~\eqref{eq:main}.  


\section{Conclusion}
We proposed a compact closed-form approximation of the average PER for coded short-packet transmission over Nakagami-$m$ fading. The error mechanism dictates the sigmoid: the maximum-of-$N$ structure of uncoded transmission yields a Gumbel limit by EVT, whereas the symmetric finite-blocklength waterfall motivates a logistic form. Interestingly, its Nakagami-$m$ average reduces to a single-term Lerch transcendent valid for arbitrary $m$, staying within about $1\%$ of the normal-approximation PER where common closed-form baselines exceed several percent, and yields a QoS-aware rate-selection rule. Extending the same approach to composite fading, such as the Fisher--Snedecor $F$ model, would require a different special-function reduction, since its SNR density lacks the gamma density's Laplace structure.

\vspace{-7pt}
\bibliographystyle{IEEEtran}
\bibliography{references}

\end{document}